\documentclass [sigconf]{acmart}
\renewcommand\footnotetextcopyrightpermission[1]{} 

\usepackage[utf8]{inputenc} 
\usepackage[T1]{fontenc}    
\usepackage{hyperref}       
\usepackage{url}            
\usepackage{booktabs}       
\usepackage{amsfonts}       
\usepackage{nicefrac}       
\usepackage{microtype}      
\usepackage{xcolor}         
\usepackage{amsmath}
\usepackage{pgfplots}
\pgfplotsset{width=14cm, height=8cm, compat=1.15}
\usepackage{graphicx}
\usepackage{tabularx}
\usepackage{caption}
\usepackage{enumitem}
\usepackage{tcolorbox}

\usepackage{amsthm}
\usepackage{algorithm}
\usepackage{algpseudocode}
\usepackage{amsmath}

\newtheorem{definition}{Definition}

\title{Explaining the “Why”: A Unified Framework for the Additive Attribution of Changes in Arbitrary Measures}

\author{Changsheng Zhou}
\email{zhouchangsheng.zcs@antgroup.com}
\affiliation{%
  \institution{Ant Group}
  \city{Hangzhou}
  \country{China}
}
  
\author{Dajun Chen}
\email{chendajun.cdj@antgroup.com}
\affiliation{%
  \institution{Ant Group}
  \city{Beijing}
  \country{China}
  }

\author{Zhitao Shen}
\email{zhitao.szt@antgroup.com}
\affiliation{%
  \institution{Ant Group}
  \city{Shanghai}
  \country{China}
  }

\author{Wei Jiang}
\email{jonny.jw@antgroup.com}
\affiliation{%
  \institution{Ant Group}
  \city{Beijing}
  \country{China}
  }

\author{Yong Li}
\email{liyong.liy@antgroup.com}
\affiliation{%
  \institution{Ant Group}
  \city{Hangzhou}
  \country{China}
  }
  \author{Peng Di}
  \authornote{corresponding author}
\email{dipeng.dp@antgroup.com}
\affiliation{%
  \institution{Ant Group}
  \city{Hangzhou}
  \country{China}
  }

\begin{abstract}

Explaining why aggregated measures change is a critical challenge in data analytics that existing systems struggle to address. While current attribution methods exist, they lack a unified solution that is simultaneously general for arbitrary measures, holistic across both data dimensions and measure composition, and rigorous in its interpretability. To bridge this gap, we introduce a principled framework that reframes attribution through the powerful lens of cooperative game theory. Our key contribution is a classification of measures based on their mathematical structure, which enables a spectrum of algorithms—from general approximations to exact, closed-form solutions—that offer a principled trade-off between generality and performance. 
We demonstrate our framework's superiority through a multi-faceted evaluation: simulations first confirm its numerical accuracy and then its generality for non-additive measures; a case study on Simpson's Paradox showcases its unique interpretability; and a final experiment proves its practical utility by significantly outperforming existing root cause analysis systems.

\end{abstract}


\begin{document}

\maketitle

\section{Introduction}

Measures, or synonymously metrics, play a vital role by providing subtle insights into business performance and aiding decision making in many areas, such as micro service monitoring~\cite{robustspot,CMMD,radice,squeeze}, advertising revenue tracking~\cite{Bhagwan2014AdtributorRD}, industrial electricity consumption analysis~\cite{ANG2015233,ANG1998489} among others. 
Typically, \emph{fundamental measures} can be aggregated to various levels over transactional data volumes that are associated with several categorical attribute dimensions, using aggregation operators such as sum, average, count, etc. 
Additionally, more complex \emph{derived measures} can be defined on top of these fundamental ones together with an arithmetic function. 
Modern OLAP tools~\cite{olap} efficiently compute aggregations but lack the ability to explain why certain values occur. 
This is insufficient for analysts, especially in anomalous cases.
Diagnosing such issues requires significant effort, including drilling down deviations across attribute dimensions or analyzing interactions between fundamental and derived measures. 
To tackle this challenge, the development of an effective \emph{attribution method} is essential to quantatively and accurately attribute the observed variations in aggregated measures.



Prior work on explaining measure variations spans multiple areas. OLAP methods identify contributing factors for multidimensional aggregate differences but assume additive measures, limiting applicability to non-additive ones~\cite{Sarawagi2001iDiff, sarawagi_explainingdifferencesinmultidimensionalaggregates, TheCascadingAnalystsAlgorithm}. 
LMDI techniques in index decomposition analyze energy consumption but are tailored to specific measure types and lack generality~\cite{ANG2015233, ANG1998489}. 
Root cause localization methods attribute and localize anomalies in applications like revenue tracking and microservices, but their attribution lacks formal additive properties and does not handle arbitrary aggregates~\cite{autoroot, CMMD, squeeze, riskloc, robustspot, idice, Bhagwan2014AdtributorRD}. 
Causal graph approaches use causal inference to explain measure changes, offering complementary insights but addressing a different problem than this work's attribution focus~\cite{radice, NEURIPS2022_c9fcd02e}.  
In summary, while these prior works have made significant strides, they collectively fall short in providing a solution that is simultaneously general for arbitrary measures, comprehensive in its ability to attribute across both sub-cubes and sub-measures, and mathematically rigorous in guaranteeing additive interpretability. These shortcomings motivate the central challenges we address.


\begin{itemize}[leftmargin=*]

\item \textbf{Generality to arbitrary measures.}  The central difficulty in attribution analysis is creating a method general enough for the complex, non-linear measures used in practice. Simple additive logic fails for derived measures like ratios (e.g., $success\_rate=success\_cnt/total\_cnt$), where the impact of one component depends on the value of another. The problem is even more acute for non-additive aggregators like $count\_distinct(\cdot)$, where the contribution of any subgroup is context-dependent and cannot be simply isolated and summed. Any general attribution framework must therefore overcome the formidable challenge of modeling these complex, interactive, and non-separable effects in a unified and systematic way.


\item \textbf{Simultaneous attribution across sub-cubes and sub-measures.} 
For complex derived measures, a deviation is driven by an interplay between which subgroup is affected (the sub-cube) and how its underlying numbers changed (the sub-measures). The core analytical challenge is that these two sources of variation interact, creating a complex attribution problem that cannot be solved by analyzing each axis in isolation. Developing a framework that can systematically untangle these joint contributions, rather than treating them separately, is a significant and unsolved hurdle.


\item \textbf{Enhanced interpretability through additivity guarantee.}
A key requirement for any trustworthy attribution is that the individual contributions must add up exactly to the total change. The central challenge in guaranteeing this additivity is that in any complex system, factors interact. The effect of one change is amplified or dampened by others, creating a shared "interaction effect." The truly hard part is to find a non-arbitrary, mathematically sound method to fully distribute this shared effect among the base factors. Without solving this allocation problem, explanations will be incomplete or misleading.

\end{itemize}

We tackle these challenges with a principled framework for attributing changes in aggregated measures, grounding our solution in cooperative game theory. The core of our approach is to treat the measure's aggregation function as the "payout" of a game and define each sub-measure $\times$ sub-cube combination as a "player". We then introduce the \emph{Shapley value} to fairly quantify contributions, providing a single, axiomatically-justified method that universally replaces measure-specific heuristics. To apply this theory in the real world, we classify measures by their mathematical structure, as this dictates the aggregation's computational complexity.

\begin{itemize}[leftmargin=*]
    \item \textbf{Generalized Additive Measures (GAMs):} This broad class includes measures that, while potentially complex and non-linear (e.g., ratios, products), are still composed of fundamental measures through well-defined, differentiable functions. They possess a structure that can be exploited for computational efficiency.
    \item \textbf{Non-Generalized Additive Measures (Non-GAMs):} This category encompasses any arbitrary measure that lacks a smooth, differentiable structure, such as those involving $count\_distinct()$ operators, conditional logic (if/else), or other black-box computations.
\end{itemize}
The classification allows us to map specific classes of measures to the most appropriate and efficient Shapley-based computation, creating a systematic and versatile attribution framework. This game-theoretic foundation, guided by our classification, elegantly and simultaneously overcomes the core challenges of generality, simultaneous attribution, and guaranteed additivity.


We conduct a comprehensive evaluation to validate our framework's accuracy, generality, interpretability, and superiority. We first establish numerical accuracy against a known ground truth via simulations on linear measures. We then verify its generality for non-additive measures, through a separate controlled experiment. Next, we demonstrate its unique explanatory power by resolving the classic Simpson’s Paradox in the Berkeley admissions dataset. Finally, we establish its practical superiority by integrating our method into five state-of-the-art root cause localization systems, consistently improving their diagnostic accuracy on a real-world anomaly dataset.




The main contributions of this paper are as follows:

\begin{itemize}[leftmargin=*]

\item We introduce a novel attribution measure approach that pioneers the use of the Shapley value, providing a single, axiom-grounded method that is general, holistic, and guarantees additive explanations for any aggregated measure.

\item We implement our framework by developing a spectrum of efficient algorithms, including classic Shapley, Aumann-Shapley, and linear approaches, that offer principled trade-offs between universal applicability and computational performance.

\item Our comprehensive experiments validate our framework's accuracy (via simulation), generality (on non-additive measures), interpretability (by resolving Simpson's Paradox), and practical superiority (by improving state-of-the-art root cause analysis systems).

\end{itemize}
\section{Related Work}
Several fields of literature have explored the topic of reasoning variations in measures.

\paragraph{Change Explanation in OLAP} Within the OLAP research domain, substantial efforts have been dedicated to developing methodologies for summarizing differences between multidimensional aggregates \cite{Sarawagi2001iDiff, sarawagi_explainingdifferencesinmultidimensionalaggregates, TheCascadingAnalystsAlgorithm}. These approaches primarily aim to identify a concise subset of contributing factors that explain the majority of observed variations across aggregates. However, a notable limitation of these methods is their inherent design focus on additive quantities, which restricts their applicability to more general, non-additive measures commonly encountered in practical scenarios.

\paragraph{Domain-Specific Decomposition Methods} In the literature on index decomposition analysis (IDA), logarithmic mean Divisia index (LMDI) decomposition methods have been increasingly used to address issues of energy consumption\cite{ANG2015233, ANG1998489,repec:arx:papers:2403.04354,SHEN2025137905}. While these methods have demonstrated effectiveness in addressing specific challenges in energy consumption analysis, their design is inherently tailored to particular classes of measures within this specialized context. This domain-specific focus restricts their broader applicability for general-purpose analysis across diverse problem domains. 

\paragraph{Multi-dimensional Root Cause Analysis (RCA)}
In the field of multi-dimensional root cause localization, numerous methods have been proposed to identify abnormalities in multi-dimensional metrics at the aggregated level by narrowing them down to specific subsets. These approaches address challenges in areas such as revenue tracking, microservice metric monitoring, and other critical business performance scenarios\cite{autoroot, CMMD, squeeze, riskloc, robustspot, idice, Bhagwan2014AdtributorRD}. Typically, these methods are structured around two core components: (1) an attribution algorithm that computes contribution factors for sub-cubes, which aligns with the problem addressed in this paper,  and (2) a pattern mining or search algorithm designed to parsimoniously identify one or several subsets that explain the primary cause of the observed difference, adhering to the principle of Occam's Razor. However, the attribution components of these methods fail to satisfy the formal fairness properties addressed in this paper, and none of them is applicable to arbitrary measure aggregates.

\paragraph{Causal Graph-Based RCA}
The literature on causal graph-based root cause analysis focuses on explaining changes in measures by incorporating causal mechanisms with other related measures\cite{radice, NEURIPS2022_c9fcd02e, 10.1145/3580305.3599392, 10.1145/3534678.3539041, causal_structure_based_rca,10.1145/3589334.3645442}. These approaches are generally structured around a two-stage framework: (1) graph construction using causal structure learning algorithms to model interrelations among various measures, and (2) application of causal inference techniques to identify the underlying causes of abnormal measures. While this class of algorithms provides valuable insights by reasoning about measure variations through causal mechanisms, they don't align with the attribution problem we address in this work.

\paragraph{Shapley Value in Explainable AI}
The most prominent and impactful recent application of the Shapley value is in Explainable AI. The SHAP framework~\cite{lundberg2017unifiedapproachinterpretingmodel} brilliantly connected various feature attribution methods (like LIME~\cite{lime}) under a unified, game-theoretic umbrella. 
Subsequent research has largely focused on creating computationally efficient approximations (e.g., TreeSHAP~\cite{lundberg2019consistentindividualizedfeatureattribution, lundberg2019explainableaitreeslocal, yang2022fasttreeshapacceleratingshap}, DeepSHAP~\cite{Chen_2022, shrikumar2017learning}, KernelSHAP~\cite{jethani2021fastshap}) and debating theoretical nuances like feature dependence and the distinction between attribution and causality~\cite{aas2021explaining, janzing2020feature}. While our work is deeply inspired by SHAP, we address a fundamentally different problem: we explain why an aggregated measure over a dataset changes, not why a model makes a single prediction. We thus adapt the game-theoretic approach from the domain of model introspection to that of data analytics and diagnostics.

\section{Preliminaries}

\begin{table*}[ht]
    \centering
    \setlength{\extrarowheight}{2pt}
    \caption{Example dataset containing online service request records. Each record is characterized by its event time,
two categorical attributes (data\_center and os\_version), and three measurable columns (request\_id, is\_success and delay)}
    \label{tab:toy data}
    \begin{tabularx}{0.8\linewidth}{>{\centering\arraybackslash}p{3cm} >{\centering\arraybackslash}X >{\centering\arraybackslash}X >{\centering\arraybackslash}X >{\centering\arraybackslash}X >{\centering\arraybackslash}X}
        \toprule
        \multicolumn{1}{c}{\textbf{$\mathcal{T}$}} & \multicolumn{2}{c}{\textbf{$\mathcal{A}$}} & \multicolumn{3}{c}{\textbf{$\mathcal{M}$}} \\
        \cmidrule(lr){1-1} \cmidrule(lr){2-3} \cmidrule(lr){4-6}
        \textbf{datetime} & \textbf{data\_center} & \textbf{os\_version} & \textbf{request\_id} & \textbf{is\_success} & \textbf{delay/ms} \\
        \midrule
        2025-01-01 00:00:00 & dc1 & v1 & 1 & 1 & 30 \\
        2025-01-01 00:00:01 & dc2 & v1 & 2 & 1 & 40 \\
        2025-01-01 00:00:02 & dc2 & v2 & 3 & 0 & \texttt{null} \\
        \bottomrule
    \end{tabularx}
\end{table*}

\begin{table*}[htbp]
    \centering
    \caption{Aggregated performance measures by data\_center. Request success rates across data centers at different times, that derived from aggregating transactional records in Table~\ref{tab:toy data}.}
    \label{tab:toy data agg}
    \begin{tabular}{l c c c c c c c}
        \toprule
        & \multicolumn{3}{c}{\textbf{10:00}} & \multicolumn{3}{c}{\textbf{10:01}} & \textbf{change} \\
        \cmidrule(lr){2-4} \cmidrule(lr){5-7} \cmidrule(lr){8-8}
        & \textbf{succ\_cnt} & \textbf{total\_cnt} & \textbf{succ\_rate} & \textbf{succ\_cnt} & \textbf{total\_cnt} & \textbf{succ\_rate} & \textbf{succ\_rate} \\
        \midrule
        \textbf{data\_center=*} & 50 & 70 & 71.43\% & 100 & 150 & 66.67\% & -4.76\% \\

        \midrule
        \textbf{data\_center=dc1} & 40 & 50 & 80\% & 45 & 50 & 95\% & +5\% \\
        \textbf{data\_center=dc2} & 10 & 20 & 50\% & 55 & 100 & 55\% & +5\% \\
        \bottomrule
    \end{tabular}
\end{table*}

\subsection{Motivation}



We begin with a motivating example to illustrate the core challenge of measure attribution. 

First, let us understand how performance data is generated. We consider a system that produces transactional records like those in the example dataset in Table~\ref{tab:toy data}. Each record has a timestamp, categorical attributes (e.g., $data\_center$), and measurable values (e.g., $is\_success$).
An OLAP system aggregates these raw records over time windows. Fundamental measures will be obtained by aggregation operators, such as $total\_cnt = COUNT(request\_id)$ and $succ\_cnt = SUM(is\_success)$.
From these, a more complex derived measure is calculated, such as our key performance indicator (KPI): $success\_rate = succ\_cnt/total\_cnt$. The components $succ\_cnt$ and $total\_cnt$ are the sub-measures of $success\_rate$.

Now, consider the aggregated data shown in Table~\ref{tab:toy data agg}, which represents the output of this process at two time steps, 10:00 and 10:01. An analyst observes an anomaly: overall $success\_rate$ has dropped by 4.76\%
A natural first step is to drill down along the $data\_center$ dimension to identify the source of the problem. However, this reveals a paradox: the $success\_rate$ within every individual data center has actually improved. The rate for $dc1$ improved by 5\%, and $dc2$ also improved by 5\%. This apparent contradiction makes root cause analysis impossible with a simple drill-down; it wrongly suggests that all components are performing better despite the aggregate decline.

The resolution to this paradox lies in understanding that a derived measure like $success\_rate$ is composed of fundamental sub-measures: $success\_rate = succ\_cnt / total\_cnt$. The true cause of the drop is not a change in departmental performance, but a change in data composition. A closer look at Table \ref{tab:toy data agg} reveals a massive shift in workload: the $total\_cnt$ for $dc2$ (the data center with a consistently lower success rate) increased five-fold, from 20 to 100. This compositional shift disproportionately weighted $dc2$'s lower rate in the overall average, overwhelming the individual performance gains in both data centers.

This example highlights the two fundamental axes of attribution that must be analyzed simultaneously:

\begin{itemize}[leftmargin=*]
\item Sub-cubes: The disaggregated segments of the data based on attributes (e.g., the data for $dc1$ or $dc2$).
\item Sub-measures: The constituent numerical components of a derived measure (e.g., $succ\_cnt$ and $total\_cnt$).
\end{itemize}

Failing to account for both can lead to misleading or entirely incorrect conclusions. This demonstrates the critical need for a comprehensive attribution framework that can systematically untangle and quantify the contributions from both dimensional changes (sub-cubes) and compositional changes (sub-measures).

\subsection{Problem Statement}

To formalize the attribution problem, we first define the core concepts of our data model.

\subsubsection{Data and Cube Structure}

We consider a standard transactional dataset $\mathcal{D}$, illustrated in Table~\ref{tab:toy data}, which contains a timestamp domain $\mathcal{T}$, a categorical attribute domain $\mathcal{A}$, and a measurable domain $\mathcal{M}$.

A \emph{data cube} is a subset of this data, defined by a predicate tuple $\nu$ over the attributes in the form of $(...,A_{i}=a_{ij}, ...)$ , where the ellipsis for attribute $A_i$ can be any corresponding attribute value $a_{ij}$ or wildcard symbol $*$ (e.g. ($data\_center=dc1$, $os\_version=*$)). For any given analysis, a higher-level hypercube $\nu$ can be partitioned into a set of disjoint sub-cubes $\{\nu'_1, \nu'_2, ..., \nu'_p\}$ by drilling down along one or more attribute dimensions. For example, the global cube `(*,*)` can be partitioned into $(dc1,*)$ and $(dc2,*)$ along the $data\_center$ dimension.


\subsubsection{Measures and Sub-Measures}
A measure $y$ is a value computed on a data cube. We are interested in derived measures, which are defined as a function of several fundamental aggregations: $y = f(m_1, m_2, ..., m_q)$. We refer to these fundamental components $\{m_1, m_2, ..., m_q\}$ as the sub-measures of $y$. For instance, for the measure $success\_rate = succ\_cnt / total\_cnt$, its sub-measures are $succ\_cnt$ and $total\_cnt$. This general formulation allows us to handle any arbitrary measure.

\subsubsection{Problem Formulation}
Now we can formally state the attribution problem. Suppose we observe that a measure $y$, calculated on a hypercube $\nu$, deviates from a reference value $y^r$ to a target value $y^t$. \textbf{Our goal} is to explain this total difference, $\Delta y = y^t - y^r$, by attributing it to its fundamental drivers.

The fundamental drivers are the combined effects of each sub-measure within each sub-cube. We aim to find a \textbf{contribution matrix} $C$ of size $p \times q$, where each element $c_{uv}$ represents the contribution of sub-measure $m_v$ from sub-cube $\nu'_u$.

This matrix must provide a complete and fair additive decomposition of the total difference, satisfying the core constraint:

\begin{equation}
\sum_{u=1}^{p} \sum_{v=1}^{q} c_{uv} = \Delta y
\end{equation}

Because $C$ is a fully additive explanation, it can be marginalized to analyze contributions at different granularities. For example, summing a row ($\sum_{v} c_{uv}$) gives the total contribution of sub-cube $\nu'_u$, while summing a column ($\sum_{u} c_{uv}$) gives the total contribution of sub-measure $m_v$ across all sub-cubes.

\subsection{Shapley Value}
\label{shapley concept}
We gently go through the concept of Shapley value~\cite{shapley:book1952} from cooperative game theory.

A coalitional game $(\sigma, N)$ can be defined by a set of players $N:=\{1,2,...,n\}$ and a set function $\sigma:2^N \rightarrow \mathbb{R}$ that measures the worth for any coalition of players $S \subseteq N$, with $\sigma(\phi)=0$ where $\phi$ is the empty set. The goal is to fairly distribute the total worth $\sigma(N)$ of the grand coalition $N$ to each player $i$. According to Shapley value, the worth of player $i$ in the game is given by

\begin{eqnarray}
\psi_i(\sigma) &=& \sum_{S\subseteq N\setminus \{i\}}\frac{|S|!(|N|-|S|-1)!}{|N|!}\left[\sigma({S\cup\{i\}})-\sigma(S)\right] \\
&=& \frac{1}{|N|}\sum_{S\subseteq N\setminus \{i\}}\binom{|N|-1}{|S|}^{-1}\left[\sigma({S\cup\{i\}})-\sigma(S)\right]
\end{eqnarray}

where $|N|$ is the cardinality of the total player set and $|S|$ is the size of a subset of players $S$. This formulation can be interpreted as follows: $\sigma({S\cup\{i\}})-\sigma(S)$ is the marginal contribution of player $i$ to the coalition after its joining, and $\frac{|S|!(|N|-|S|-1)!}{|N|!}$ is the weight of coalition $S$. Then by averaging the worth of player $i$ over all possible subset $S$ that excludes player $i$ including the empty subset $\phi$, we arrive at the Shapley value of $i$ with removal of randomness of $S$.
Shapley value has another equivalent formulation:
\begin{equation}
\psi_i(\sigma)=\frac{1}{|N|!}\sum_{r\in\mathcal{R}}[\sigma(P_i^r\cup{i})-\sigma(P_i^r)]
\end{equation}
where $\mathcal{R}:N \rightarrow N$ is all possible permutations of the players and $P_i^r$ is the coalition that precedes player $i$ for a specific permutation $r$. Averaging over all $|N|!$ permutations gives us the exact calculation of Shapley value.

Shapley value has been proved to be the unique additive solution to credit distribution that satisfies the following desirable properties to guarantee fairness\cite{lundberg2017unifiedapproachinterpretingmodel}.

\begin{itemize}[leftmargin=*]
    
\item \textbf{Completeness} The Shapley values of all players sum up to the worth of the grand coalition. That is, $\sum_{i=1}^n\psi_i(\sigma)=\sigma(N)$.

\item \textbf{Dummy} The dummy players that make no contribution to any coalitions get zero Shapley values. That is, if for any coalition $S \subseteq N\setminus\{i\}$, $\sigma(S\cup{i})=\sigma(S)$, then $\psi_{i}(\sigma)=0$.

\item \textbf{Symmetry} If two players contribute the same to any coalition that excludes them, they have equal Shapley values. That is, if for any coalition $S \subseteq N\setminus\{i,j\}$, $\sigma(S\cup{i})=\sigma(S\cup j)$, then $\psi_{i}(\sigma)=\psi_{j}(\sigma)$.

\item \textbf{Linearity} If a set function $\sigma$ is the sum of two set functions $\sigma_1$ and $\sigma_2$, then the Shapley value of any player of $\sigma$ is the sum of that of $\sigma_1$ and that of $\sigma_2$. That is, for any $\sigma_1$ and $\sigma_2$, $\psi_i(\sigma_1+\sigma_2)=\psi_i(\sigma_1)+\psi_i(\sigma_2)$.

\end{itemize}
\section{Methodology}
We aim to introduce a general attribution framework that is applicable to arbitrary measures. However, generalized additivity of measures is a desirable property that enables a broader range of approaches. To this end, we first present methods tailored for generalized additive measures and subsequently we discuss non-generalized additive cases.

\subsection{Generalized Additive Measures}

To systematically analyze contributions, our first step is to structure the disaggregated data into a unified \textbf{observation matrix}. 
For a given hypercube $\nu$ partitioned into $p$ sub-cubes $\{\nu'_1, \dots, \nu'_p\}$ and a measure $y$ composed of $q$ sub-measures $\{m_1, \dots, m_q\}$, we construct a matrix $X \in \mathbb{R}^{p \times q}$. 
Each element $x_{uv}$ in this matrix represents the value of the $v$-th sub-measure ($m_v$) calculated on the $u$-th sub-cube ($\nu'_u$).

The key to simplifying our analysis lies in a property of the sub-measures, which we formally define as additivity.

\begin{definition}[Additivity]
A measure $m_v$ is said to be \textbf{additive} if its value on a parent cube is the sum of its values on its constituent sub-cubes. Formally, for any drilling dimension $A_j$:
\begin{equation}
    h_v\left(\mathcal{D}_\mathcal{M}^{\Gamma^{\tau}({...,A_j=*,...})}\right) = \sum_{k=1}^{|A_j|} h_v\left(\mathcal{D}_\mathcal{M}^{\Gamma^{\tau}({...,A_j=a_{jk},...})}\right)
\end{equation}
\end{definition}

This property is crucial because if a sub-measure $m_v$ is additive, its total value across the hypercube $\nu$ is simply the sum of the $v$-th column of our observation matrix: $\sum_{u=1}^p x_{uv}$. This allows us to define a special, broad class of measures that are particularly amenable to our analysis.

\begin{definition}[Generalized Additive Measure (GAM)]
A measure $y$ is a \textbf{Generalized Additive Measure (GAM)} if all of its sub-measures $\{m_1, \dots, m_q\}$ are additive.
\end{definition}

For any GAM, the total measure $y$ can be expressed as a direct function of the column sums of the observation matrix $X$:
\begin{equation}
y = f\left(\sum_{u=1}^p x_{u,1}, \sum_{u=1}^p x_{u,2}, \dots, \sum_{u=1}^p x_{u,q}\right) \equiv g(X)
\end{equation}

\subsubsection{Problem Reformulation}
With this structure, our attribution problem is significantly clarified. We obtain an observation matrix $X^t$ for the target time step (the \textbf{explicand}) and another, $X^r$, for the reference time step. The problem then reduces to explaining the difference $\Delta y = y^t - y^r$ by attributing it to the changes in the individual elements of the matrix, as the overall value is now a direct function of these matrix inputs:
\begin{equation}
    \Delta y = g(X^t) - g(X^r)
\end{equation}

\subsubsection{Attribution Algorithms}
Our framework provides a spectrum of algorithms to compute the contribution matrix $C$. These methods offer principled trade-offs between universal applicability and computational efficiency.

\paragraph{Classic Shapley Value: A Universal Approach}
This is the most fundamental approach, applicable to any measure, regardless of its mathematical form. The intuition is that the contribution of a factor $x_{uv}$ depends on the order in which it is considered relative to others. To ensure fairness and eliminate this arbitrariness, the Shapley value averages the marginal contribution over all possible permutations.

We formalize this by defining a coalitional game $\mathcal{G}=(\sigma,X)$. We use a boolean mask matrix $Z \in \{0,1\}^{p \times q}$ to represent a coalition, where $z_{uv}=1$ indicates that the value $x_{uv}^t$ from the explicand is used, and $z_{uv}=0$ indicates the value $x_{uv}^r$ from the reference is used. The set function $\sigma(Z)$ defines the total worth of a coalition:
\begin{equation}
    \sigma(Z) = g(X^t \odot Z + X^r \odot \overline{Z}) - g(X^r)
\end{equation} 
where $\overline{Z}$ is the complement of $Z$ and $\odot$ is the Hadamard product. The goal is to solve the game to find the contribution matrix $C$:
\begin{equation}
    C = \zeta(\mathcal{G}, X^t, X^r) \label{eq:coalitional_game}
\end{equation}


\textbf{Efficiency:} The exact calculation of the Shapley value is computationally prohibitive, with a time complexity of $O(2^{pq}\cdot pq)$. While efficient approximation methods like KernelSHAP ($O(k \cdot (pq)^2)$, where $k$ is the sample size of subsets of game players) have been developed to make this practical for general applications~\cite{lundberg2017unifiedapproachinterpretingmodel, chen2022algorithmsestimateshapleyvalue, jethani2021fastshap}. 

In many common use cases, the full complexity is unnecessary because the analytical question allows attribution to be performed along a single dimension. This ability to analyze dimensions independently is key to our framework's practical performance. For example, when attributing changes to the $q$ sub-measures, the complexity is only $O(q \cdot 2^q)$, which is often tractable for exact calculation. Similarly, when attributing to the $p$ data partitions, we can use approximations with a more manageable complexity of $O(k \cdot p^2)$. This layered approach allows for exact, rapid computation in common scenarios while still supporting full, holistic analysis via efficient approximation when required.



\paragraph{Aumann-Shapley Value: An Efficient Approach for Differentiable Measures}
For the broad class of measures where the function $g(X)$ is differentiable, we can use the Aumann-Shapley value, which is equivalent to Integrated Gradients~\cite{integratedgradients, themanyshapleyvaluesformodelexplanation}. This is often far more efficient and defines the contribution of $x_{uv}$ as the path integral of its gradient along a straight line from the reference $X^r$ to the explicand $X^t$:
\begin{equation}
    c_{uv} := (x_{uv}^t - x_{uv}^r) \int_{\alpha=0}^{1} \frac{\partial g\left(X^r + \alpha(X^t - X^r)\right)}{\partial x_{uv}} d\alpha \label{eq:ig}
\end{equation}
The resulting integral does not always have a simple form, but it can be handled in two ways:
\begin{itemize}[leftmargin=*]
    \item \textbf{Riemann Integral Approximation.} The integral can be efficiently approximated by summing the gradients at discrete steps along the path:
    \begin{equation}
        \widehat{c}_{uv} = \frac{(x_{uv}^t-x_{uv}^r)}{m} \sum_{k=1}^{m}\frac{\partial g\left(X^r+\frac{k}{m}(X^t-X^r)\right)}{\partial x_{uv}}
    \end{equation}
    where $m$ is the number of approximation steps (typically 100-1000).

    \item \textbf{Closed-Form Solution.} For certain important functions, closed-form solutions exist. For the common ratio measure $y = m_1/m_2$, we derive the exact solution by solving Eq.~\ref{eq:ig}. Denoting $x_i^t=\sum_u x_{ui}^t$ and $x_i^r=\sum_u x_{ui}^r$, the contributions are:
    \begin{eqnarray}
        c_{u1} = (x_{u1}^t-x_{u1}^r)\frac{1}{x_2^t-x_2^r}\ln\frac{x_2^t}{x_2^r} \\
        c_{u2} = \left(x_{u2}^t-x_{u2}^r\right)\left(\frac{-x_1^r}{x_2^rx_2^t}-\frac{x_1^t-x_1^r}{(x_2^t-x_2^r)^2}\left(\ln\frac{x_2^t}{x_2^r}+\frac{x_2^r}{x_2^t}-1\right)\right)
    \end{eqnarray}
    In the edge case where $x_2^t = x_2^r$, the limits are well-defined and can be calculated via Taylor expansion.
\begin{eqnarray}
\lim_{x_2^t \to x_2^r} c_{u1} = (x_{u1}^t-x_{u1}^r) \frac{1}{x_2^r}\\
\lim_{x_2^t \to x_2^r} c_{u2}=(x_{u2}^t-x_{u2}^r)\frac{-x_1^t}{x_2^r x_2^t}
\end{eqnarray}
\end{itemize}
\textbf{Efficiency:} The Riemann approximation has complexity $O(m \cdot pq)$. When a closed-form solution exists, the complexity reduces to constant time with fixed $p$ and $q$.

\paragraph{Linear Approach: An Instantaneous Solution for Linear Measures}
For the simplest case where the measure is a linear combination of its sub-measures, $y = w_0 + \sum_{v=1}^q w_v \cdot m_v = w_0 + \mathbf{j}X\mathbf{w}$, attribution is trivial and instantaneous. The contribution of each factor has an exact, closed-form solution:
\begin{equation}
    c_{uv} = w_v \cdot (x_{uv}^t - x_{uv}^r)
\end{equation}
\textbf{Efficiency:} The linear approach has constant time complexity with fixed $p$ and $q$.

\subsubsection{Discussion on Reference Data}
A crucial component for all the above algorithms is the definition of the reference data, $X^r$. The choice of reference frames the attribution question being asked, and our framework supports two primary modes:
\begin{itemize}[leftmargin=*]
    \item \textbf{Baseline Shapley.} This method compares the target explicand $X^t$ against a single, fixed reference point $X^r$. It is highly practical for common business scenarios like month-on-month or year-on-year comparisons, where the previous period serves as a natural and intuitive baseline.

    \item \textbf{Expected Shapley.} This provides a more robust comparison against a reference distribution $p(X)$, which is useful when a single reference point is noisy or arbitrary. The attribution is the average contribution over many possible baselines sampled from $p(X)$. The contribution matrix is given by \begin{equation}
    \label{eq:approx expected shapley}
    C=\mathbb{E}_{X^\prime\sim p(X)}\left[\zeta(\mathcal{G},X^t,X')\right]\end{equation}, which can be approximated by averaging over a sample set $\mathcal{X}$:
    \begin{equation}    \label{eq:approx expected shapley hat}
        \hat{C} = \frac{1}{|\mathcal{X}|}\sum_{X' \in \mathcal{X}} \zeta(\mathcal{G}, X^t, X')
    \end{equation}
    The distribution $p(X)$ can be an empirical distribution from normal time steps or derived from a more sophisticated time-series model to handle trends or seasonality.
\end{itemize}

\subsection{Non-Generalized Additive Measures}
\label{non gam}

In the preceding sections, we focused on Generalized Additive Measures (GAMs), where pre-aggregation allows for highly efficient attribution. However, many real-world measures are not additive. For these Non-GAMs, the total value on a hypercube is not a simple function of its sub-cube values, rendering the pre-aggregated matrix $X$ invalid. This means the efficient Aumann-Shapley and Linear approaches are no longer viable~\footnote{A special case is the $avg(\cdot)$ aggregator, a non-generalized additive measure by itself. However, it can be decomposed into $avg(\cdot)=sum(\cdot)/count(\cdot)$, which is generalized additive, still allowing for Aumann-Shapley value approach.}.

Nevertheless, the classic Shapley value approach remains applicable due to its model-agnostic nature. The core logic is the same, but we must adapt the coalitional game to operate at a deeper level.

\subsubsection{The Adapted Game: Attributing from Raw Data}
For GAMs, we played the game by substituting pre-aggregated values $\{x_{uv}\}$ in our observation matrix. For Non-GAMs, we must play the game by substituting the \textbf{raw data records} from $\mathcal{D}$ directly before each aggregation step.

To formalize this, we re-employ the coalition mask matrix $Z \in \{0,1\}^{p \times q}$. For each evaluation of the set function, we construct a temporary, mixed dataset for each sub-measure $m_i$. This dataset, $D^{Z,i}$, is formed by selecting records for each sub-cube based on the coalition mask:
\begin{equation}
    D^{Z,i} = \left( \bigcup_{z_{j,i}=1}\mathcal{D}^{\Gamma^t(\nu'_j)} \right)  \bigcup  \left( \bigcup_{z_{j,i}=0}\mathcal{D}^{\Gamma^r(\nu'_j)} \right)
\end{equation}
In simple terms, if a factor $(u,v)$ is in the coalition ($z_{uv}=1$), we use the raw data from the target time step $t$ for sub-cube $\nu'_u$ when calculating sub-measure $m_v$; otherwise, we use data from the reference time step $r$. The sub-measure's value is then $m_v = h_v(D^{Z,v}_\mathcal{M})$.

The set function for the coalitional game is then redefined to operate on these dynamically constructed datasets:
\begin{equation}
    \sigma(Z) = f\left(\dots, h_i\left(D^{Z,i}_\mathcal{M}\right), \dots\right) - f\left(\dots, h_i\left(D^{O,i}_\mathcal{M}\right), \dots\right)
\end{equation}
where $O$ is the all-zero matrix representing the empty coalition (i.e., the reference state).

As a result, the coalitional game can be defined as $\mathcal{G}=(\delta, \mathcal{D})$. This reformulation of the coalitional game is quite straightforward in that \textbf{it only replaces the substitution of the pre-aggregated ${x_{uv}}'s$ with the substitution of raw records from $\mathcal{D}$ directly}, since pre-aggregation for non-generalized additive measures is infeasible.

\subsubsection{The Critical Trade-Off: Generality vs. Efficiency}
This adaptation makes our framework \textbf{universally applicable} to any measure. However, this generality comes at a significant computational cost. While the GAM approach benefits enormously from pre-aggregation, the Non-GAM approach must re-aggregate from raw transactional data every time the set function $\sigma(Z)$ is evaluated. For sampling-based Shapley approximations that require thousands of evaluations, this introduces substantial overhead.

Despite this overhead, the Non-GAMs approach remains feasible and valuable in those latency-insensitive scenarios. This performance limitation also makes the method potentially intractable for very large datasets and highlights a critical area for future research and optimization. For GAMs, it is always advisable to use the pre-aggregation method for its significant efficiency gains.

\section{Experiment}

To validate our framework's accuracy, generality, interpretability, and practical superiority, we conduct four experiments, each designed to answer a key research question:
\begin{itemize}[leftmargin=*]
    \item \textbf{RQ1: Accuracy.} Can our method accurately quantify contributions?
    \item \textbf{RQ2: Generality.} Is our method applicable to complex Non-Generalized Additive Measures (Non-GAMs)?
    \item \textbf{RQ3: Interpretability.} Can our method produce insightful and interpretable explanations for complex, real-world phenomena like Simpson's Paradox?
    \item \textbf{RQ4: Superiority.} Does our method outperform state-of-the-art techniques in a practical application like root cause localization?
\end{itemize}

\subsection{RQ1: Accuracy in a Controlled Setting}
\label{exp:simulaiton_linear}

\paragraph{Objective}
To validate the numerical correctness of our framework, we conduct a simulation with a linear measure where the ground truth attribution is known, allowing us to precisely measure our method's accuracy.

\paragraph{Setup}
We simulate a linear measure $y = \sum_{i=1}^{q} m_i$ over a hypercube with $p$ sub-cubes. For this setting, the ground truth contribution of a change in factor $x_{ij}$ is simply the change itself. A single simulation proceeds as follows:
\begin{enumerate}[leftmargin=*]
    \item We uniformly randomly sample $q$ from $\{1,2,3,4,5\}$ and $p$ from between 10 and 100. 
    \item We sample all means ${\mu_{ij}}$ from a uniform distribution $\mathcal{U}(0,10)$. 
    \item We sample fault indicators ${\beta_{ij}}$ from a Bernoulli distribution $\mathcal{B}(0.5)$ and fault magnitudes ${\lambda_{ij}}$ from a uniform distribution $\mathcal{U}(0,10)$.
    \item We obtain an anomalous data point $\hat{X}$ where $\hat{x}_{ij}=\mu_{ij}+\beta_{ij} \cdot \lambda_{ij}$. The ground truth contribution is thus $\beta_{ij} \cdot \lambda_{ij}$.
    \item We sample a reference dataset $\mathcal{X}$ of size $N$ from the normal distribution $\mathcal{N}(\mu_{ij},1)$. 
    \item We approximate the attribution matrix $\hat{C}$ using the Expected Shapley method in Eq.~\ref{eq:approx expected shapley} and Eq.~\ref{eq:approx expected shapley hat}.
    \item We measure the error $e$ using Mean Absolute Scaled Error (MASE) against the ground truth.
        \begin{equation}
e=\frac{\sum_{i,j}{|\hat{c}_{ij} - \beta_{ij} \cdot \lambda_{ij}|}}{\sum_{ij}|\beta_{ij} \cdot \lambda_{ij}|}.
    \end{equation}
\end{enumerate}

We repeat this procedure 100 times for each reference sample size $N \in \{100, 200, \dots, 1000\}$.

\begin{figure}[tbp]
  \centering
  \begin{tikzpicture}
    \begin{axis}[
      xlabel={Sample Size N}, 
      ylabel={Mean Absolute Scaled Error \%}, 
      legend pos=north west, 
      width=0.5\textwidth, 
      height=0.3\textwidth, 
      ]
      y tick label style={/pgf/number format/.cd, fixed, precision=2}, 
      scaled y ticks=false, 

      \addplot [
        dashed, 
        mark=asterisk, 
        error bars/.cd, 
        y dir=both, 
        y explicit,
        error bar style={line width=1pt, solid},
        error mark options={line width=1pt, mark size=2pt, rotate=90}
      ] table [x=x, y=y, y error=y-err] {%
        x y y-err
        100 3.274113080852917 0.5823280951269653
        200 2.306187864249573 0.4906295725997777
        300 1.8996735232386767 0.29496226176381154
        400 1.6301238342268293 0.27349810098473647
        500 1.4687502190321328 0.25653748329985677
        600 1.2985656218278092 0.20590590645311398
        700 1.2341698979608852 0.19825361603958637
        800 1.1613162603544152 0.16216904079608288
        900 1.0748277115284123 0.1698268346666751
        1000 1.0367355148111056 0.17731664458228926
      };
    \end{axis}
  \end{tikzpicture}
  \caption{Mean absolute scaled error (MASE) against sample size. Error bars indicate the standard error of MASE across multiple runs of simulation.}
  \label{fig:simulation} 
\end{figure}
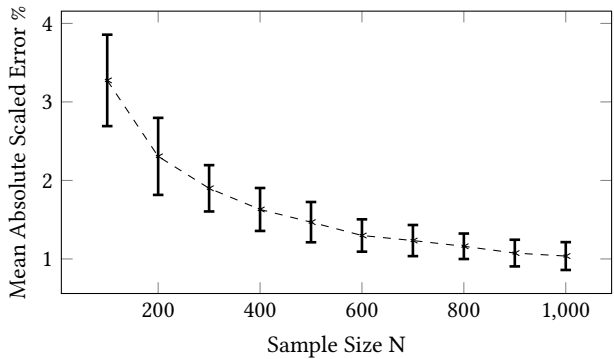

\paragraph{Results and Analysis}
Figure~\ref{fig:simulation} shows that the MASE consistently decreases as the reference sample size $N$ grows, asymptotically approaching zero. This convergence confirms that our Expected Shapley implementation is numerically sound and that with a sufficient reference sample, it accurately estimates the true contributions. 
The observed trend of diminishing returns, particularly for sample sizes $N > 500$ in our simulation, provides a practical guideline for balancing computational cost with accuracy.

\noindent
\begin{tcolorbox}[size=title, opacityfill=0.1, nobeforeafter]
\textbf{Answer to RQ1: }\textit{ As is shown by the experiment results, our method is capable of giving accurate quantified contributions  with the measured errors converging to near-zero. }
\end{tcolorbox}

\subsection{RQ2: Generality for Non-GAMs}
\label{exp:simulation_non_gam}

\paragraph{Objective}
To test our framework's generality, we conduct a simulation for a complex Non-GAM scenario: attributing changes in Daily Active Users (DAU), a measure based on the non-additive $DAU=count\_distinct(user\_id)$ operation.

\paragraph{Setup}
We simulate a website with 5 web pages and 10,000 users. A single simulation proceeds as follows:
\begin{enumerate}[leftmargin=*]
    \item We uniformly randomly sample the number of faulty pages $w_f$ from \{1,2,3\}.
    \item We sample a set of faulty web pages $\mathcal{W}_f$ of size $w_f$ from the total cardinality $\mathcal{W}$ with no replacement to act as the ground truth.
    \item We obtain a faulty user-page viewing probability matrix $P_f$ by reducing the view probability of faulty pages by a decaying factor $\lambda$.
    \item We obtain anomalous user-page viewing facts $V_f:\{0,1\}^{u \times w}$ by sampling from the Bernoulli $\mathcal{B}(P_f)$, where 1 indicates page view happened and 0 means not.
    \item We sample a set of 10 normal reference viewing facts $\{V_1, \dots, V_{10}\}$ from the original probability matrix, initialized with a constant value of 0.05.
    \item We use our Non-GAM attribution method to compute the contribution of each web page to the overall drop in DAU.
    \item We measure accuracy as the proportion of correctly identified faulty pages within the top-$w_f$ contributors.
\end{enumerate}

We repeat this procedure 20 times for each decay factor $\lambda \in \{0.1, 0.2, \dots, 1.0\}$.

\begin{figure}[tbp]
  \centering
  \begin{tikzpicture}
    \begin{axis}[
      ymin=0, ymax=1.2,    
      xlabel={Decaying Factor $\lambda$}, 
      ylabel={Accuracy}, 
      legend pos=north west, 
      width=0.5\textwidth, 
      height=0.3\textwidth, 
      ]
      y tick label style={/pgf/number format/.cd, fixed, precision=2}, 
      scaled y ticks=false, 

      \addplot [
        dashed, 
        mark=asterisk, 
        error bars/.cd, 
        y dir=both, 
        y explicit,
        error bar style={line width=1pt, solid},
        error mark options={line width=1pt, mark size=2pt, rotate=90}
      ] table [x=x, y=y, y error=y-err] {%
        x y y-err
        1.0   1.0   0.0
0.9   1.0   0.0
0.8   1.0   0.0
0.7   1.0   0.0
0.6   1.0   0.0
0.5   1.0   0.0
0.4   1.0   0.0
0.3   1.0   0.0
0.2   1.0   0.0
0.1   0.875   0.19632031648982912
      };
    \end{axis}
  \end{tikzpicture}
  \caption{ Accuracy against decay factor. Error bars indicate the standard error of accuracy across multiple runs of simulations. }
  \label{fig:simulation2} 
\end{figure}
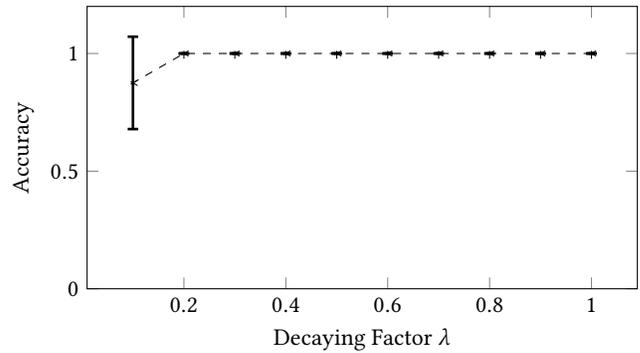

\paragraph{Results and Analysis}
Figure~\ref{fig:simulation2} shows that our method achieves perfect accuracy for faults with a severity ($\lambda$) of 20\% or more. The slight drop in accuracy for very subtle faults (10\% severity) is expected, as the true signal becomes difficult to distinguish from the random noise inherent in user behavior sampling. This result is significant because it demonstrates our framework's ability to handle the complex, combinatorial interactions present in a $count\_distinct(\cdot)$ operation, a task for which derivative, based or simple difference methods are fundamentally unsuitable.

\noindent
\begin{tcolorbox}[size=title, opacityfill=0.1, nobeforeafter]
\textbf{Answer to RQ2: }\textit{ Experiments show that our method achieves satisfactory accuracy in attributing the DAU measure, which is aggregated by the $count\_distinct(\cdot)$ operator, justifying its generality to non-GAMs. }
\end{tcolorbox}

\begin{table*}[htbp]
    \centering
    \caption{Application and admission data of UC Berkeley. An analysis of UC Berkeley’s application and admission data reveals a notable gender disparity at the aggregate level: male applicants exhibit a significantly higher admission rate than their female counterparts. 
    }
    \label{tab:berkeley data}
    \begin{tabular}{ccccccc}
        \toprule
        & \multicolumn{3}{c}{\textbf{Male}} & \multicolumn{3}{c}{\textbf{Female}} \\
        \cmidrule(lr){2-4} \cmidrule(lr){5-7}
        \textbf{Department} & \textbf{Applicants} & \textbf{Admitted} & \textbf{Admitted Rate} & \textbf{Applicants} & \textbf{Admitted} & \textbf{Admitted Rate} \\
        \midrule
        A & 825 & 512 & 62.06\% & 108 & 89 & \textbf{82.41\%} \\
        B & 560 & 353 & 63.04\% & 25 & 17 & \textbf{68.00\%} \\
        C & 325 & 120 & \textbf{36.92\%} & 593 & 201 & 33.90\% \\
        D & 417 & 138 & 33.09\% & 375 & 131 & \textbf{34.93\%} \\
        E & 191 & 53 & \textbf{27.75\%} & 393 & 94 & 23.92\% \\
        F & 373 & 22 & 5.90\% & 341 & 25 & \textbf{7.33\%} \\
        \midrule
        \textbf{Total} & 2691 & 1198 & \textbf{44.52\%} & 1835 & 557 & 30.35\% \\
        \bottomrule
    \end{tabular}
\end{table*}

\begin{table}[htbp]
    \centering
    \setlength{\extrarowheight}{2pt}
    \caption{Contribution matrix distributing difference of admission rate between males and females. 
    }
    \label{tab:berkeley contribution}

    \begin{tabular}{cccc}
        \toprule
        \textbf{Department} & \textbf{Applicants} & \textbf{Admitted} & \textbf{Total} \\
        \midrule
        A & +12.15\% & -18.92\% & \textbf{-6.77\%} \\
        B & +9.07\% & -15.03\% & \textbf{-5.96\%} \\
        E & -3.42\% & +1.83\% & \textbf{-1.59\%} \\
        C & -4.54\% & +3.62\% & \textbf{-0.87\%} \\
        D & +0.71\% & -0.31\% & +0.40\% \\
        F & +0.54\% & +0.13\% & +0.63\% \\
        \midrule
        \textbf{Total} & +14.51\% & -28.68\% & \textbf{-14.17\%} \\
        \bottomrule
    \end{tabular}
\end{table}

\subsection{RQ3: Interpretability on Simpson's Paradox}
\label{exp:berkely}

\paragraph{Objective}
To assess our method's interpretability and explanatory power, we apply it to a classic and challenging real-world scenario: the 1973 Berkeley graduate admissions data~\cite{berkeleydata}. This dataset is a famous example of Simpson's Paradox, a statistical phenomenon where a trend appears in several different groups of data but disappears or reverses when these groups are combined. It serves as a perfect litmus test for any attribution method, as simple drill-down analysis is not just insufficient, but actively misleading.

\paragraph{The Paradox in the Data}
The core of the paradox is shown in Table~\ref{tab:berkeley data}. At the aggregate level, the data suggests a significant gender bias against female applicants, whose overall admission rate is 14.17\% lower than that of male applicants (30.35\% vs. 44.52\%). This aggregate view would suggest a systemic problem. However, a disaggregated view at the department level tells a completely different story. In four of the six largest departments (A, B, D, F), female applicants actually had a \emph{higher} admission rate than their male counterparts. This contradiction makes it impossible to pinpoint the source of the aggregate disparity using simple analysis.

\paragraph{Attribution Setup}
We use our framework to provide a definitive, quantitative explanation. We set the explicand as the female applicant data and the reference as the male applicant data. The goal is to attribute the total difference of -14.17\% in the derived measure, $admitted\_rate=admitted/applicants$, across two dimensions: the six departments and the two sub-measures ($applicants$ and $admitted$). We use the Aumann-Shapley closed-form solution for ratios to generate the contribution matrix shown in Table~\ref{tab:berkeley contribution}.

\paragraph{Holistic Interpretation: Unraveling the Paradox}
The contribution matrix in Table~\ref{tab:berkeley contribution} provides a complete numerical explanation. We can interpret it at three levels to fully resolve the paradox.

\begin{itemize}[leftmargin=*]
\item \textbf{Level 1: Identifying High-Impact Departments:} The -14.17\% gap largely stems from two departments: Department A (-6.77\%) and Department B (-5.96\%). At a glance, one might mistakenly label these departments as the most biased.

\item \textbf{Level 2: Breaking Down Departmental Contributions:} A deeper analysis shows Department A's -6.77\% is composed of a +12.15\% from $applicants$ (fewer women applying helped women’s rate) and a -18.92\% from $admitted$ (far more men were admitted, dragging the rate down). This pattern, also seen in Department B, demonstrates the gap isn’t due to departmental bias but rather to large, high-admission departments with male-dominated applicant pools.

\item \textbf{Level 3: System-Wide Explanation:} The total contributions reveal Simpson’s Paradox: the $applicants$ sub-measure adds +14.51\%, while the $admitted$ sub-measure subtracts -28.68\%. The gap arises not from departmental bias, but from a compositional shift women and men applied to different departments at different rates. This analysis reveals that systemic application patterns, not performance, are the primary cause of the disparity.
\end{itemize}

\noindent
\begin{tcolorbox}[size=title, opacityfill=0.1, nobeforeafter]
\textbf{Answer to RQ3: }\textit{ Our method enables comprehensively interpreting complex real-world phenomena by systematically distributing observed measure deviations to all contributing factors. }
\end{tcolorbox}

\begin{table*}[!htbp]
    \centering
    \setlength{\extrarowheight}{2pt}
    \caption{F1-scores of root cause analysis algorithms in their original form versus when enhanced by our Shapley-based framework (Permutation, Kernel, Aumann), with optimal improvement(rate) listed below.
}
    \label{tab:multidimensional rca}
    \begin{tabular}{cccccc}
        \toprule
        \textbf{Method} & \textbf{R-Adtributor} & \textbf{RobustSpot} & \textbf{Hotspot} & \textbf{Squeeze} & \textbf{AutoRoot} \\
        \midrule
        Original & 0.1041 & 0.4034 & 0.0444 & 0.0434 & 0.2606 \\
        
        Permutation & 0.1080 & 0.4143 & 0.3250 & 0.1913 & 0.4483 \\
        
        Kernel & \textbf{0.1199} & 0.4116 & 0.3536 & \textbf{0.2724} & \textbf{0.5090} \\

        Aumann & 0.1092 & \textbf{0.4150} & \textbf{0.3731} & 0.2444 & 0.5073 \\
        \midrule

        Optimal Improvement(rate) & 0.0158 (15.2\%) & 0.0116 (2.9\%) & 0.3287 (740.3\%) & 0.2290 (527.6\%)& 0.2484 (95.3\%)\\ 
        \bottomrule
    \end{tabular}
\end{table*}

\subsection{RQ4: Practical Superiority in RCA}
\label{exp:rca}

\paragraph{Objective}
The ultimate test of an attribution method is its ability to improve real-world diagnostic applications. In this experiment, we aim to demonstrate that our framework is not just a theoretical construct but a practically superior component that can enhance existing state-of-the-art systems. We hypothesize that by replacing the often heuristic-based attribution modules of existing RCA systems with our principled approach, we can significantly improve their accuracy in identifying the correct root causes of anomalies.

\paragraph{Experiment Setup}
We use a public, real-world dataset from a leading global online video service provider, which contains \textbf{135 real-world anomalous cases} that were manually investigated and validated by experienced operators~\cite{robustspot}. The key performance measure is the \textbf{Stalling Ratio (SR)}, a non-trivial derived measure calculated as:
\[
    \text{SR} = \frac{\text{Number of Viewers Experiencing Stalling (SV)}}{\text{Total Online Viewers (OV)}}
\]
The data is structured across five attribute dimensions: CDN, Bitrate, Device Type, P2P, and ISP.

We evaluate five state-of-the-art multi-dimensional RCA algorithms from their open-source implementations~\cite{rcaframeworks}. These algorithms fall into two categories based on their native attribution logic:
\begin{itemize}[leftmargin=*]
    \item \textbf{Potential Score-Based (Heuristic):} \textit{Hotspot}~\cite{hotspot}, \textit{Squeeze}~\cite{squeeze}, \textit{AutoRoot}~\cite{autoroot}. These methods estimate influence using heuristics based on a "ripple effect" assumption.
    \item \textbf{Explanatory Power-Based (Approximation):} \textit{R-Adtributor}~\cite{radtributor}, \textit{RobustSpot}~\cite{robustspot}. These methods quantify impact by measuring the change after replacing a single sub-cube's value, i.e. a one-step approximation of marginal contribution.
\end{itemize}
For each algorithm, we compare its original F1-score against versions where we replace its attribution module with ours, using three variants: Permutation Shapley, KernelSHAP, and the closed-form Aumann-Shapley solution.

\paragraph{Results and Analysis}
The results, presented in Table~\ref{tab:multidimensional rca}, show a consistent and often dramatic improvement in F1-score across all five baseline algorithms when our attribution framework is used. A deeper analysis reveals two distinct patterns of improvement.

\begin{itemize}[leftmargin=*]
    \item \textbf{Massive Improvement over Heuristic Methods.}
    The most significant gains are seen with the potential score-based methods ($Hotspot$, $Squeeze$, $AutoRoot$). For example, Hotspot's F1-score leaps from a very low 0.0444 to 0.3731, an improvement of over 700\%. This is because their native attribution logic is based on heuristics that do not correctly model the non-linear interactions inherent in a ratio metric like SR. These heuristics provide a weak, often misleading signal to their downstream search algorithms. Our framework's principled attribution scores accurately capture sub-cube influences, enabling effective search and boosting diagnostic accuracy.

    \item \textbf{Refined Improvement over Approximation-Based Methods.}
    For the explanatory power-based methods (\textit{R-Adtributor}, \textit{RobustSpot}), the improvement is more modest but still consistent and significant. This is because their attribution logic is already a reasonable, albeit flawed, approximation of our method. Calculating the explanatory power of a single sub-cube is equivalent to calculating its marginal contribution in just \emph{one specific permutation} (where it is considered last). This makes it sensitive to ordering effects and unable to fully capture complex interactions. Our Shapley-based methods provide a more robust and refined signal by averaging over all possible permutations, effectively de-biasing their one-step approximation. This superior signal leads to more accurate root cause identification, even when the baseline is already strong.

    \item \textbf{Validation of the Algorithm Spectrum.}
    Comparing our own variants, the Aumann-Shapley closed-form solution consistently delivers top-tier performance. This is expected, as it provides the exact, ground-truth contribution for a differentiable measure like SR, free from any sampling error. This result validates our core proposal: when an efficient, exact method is available for a given measure, it should be used. The strong performance of the Kernel and Permutation approximations further demonstrates that even our general-purpose methods are superior to the specialized heuristics found in existing systems.
\end{itemize}

\noindent
\begin{tcolorbox}[size=title, opacityfill=0.1, nobeforeafter]
\textbf{Answer to RQ4: }\textit{ 
The results show our proposed framework significantly improves five RCA systems' F1-score, establishing its practical superiority.
}
\end{tcolorbox}
\section{Conclusion and Future Work}
\label{section:conclusion}

\paragraph{Discussion and Conclusion}
Existing attribution methods are often ad-hoc, measure-specific, or lack formal fairness guarantees. In this work, we introduced a unified framework grounded in cooperative game theory that overcomes these limitations. By leveraging the Shapley value, we provide a principled and holistic method for attributing changes in arbitrary aggregated measures across both data dimensions and compositional sub-measures. Our experiments validated that this approach is not only numerically accurate and general-purpose but also highly interpretable, capable of resolving complex phenomena like Simpson's Paradox. Crucially, we demonstrated its practical superiority by showing it significantly improves the diagnostic accuracy of state-of-the-art root cause analysis systems. This work lays a rigorous foundation for a new generation of truly explanatory data analysis systems.

\paragraph{Limitations and Future Work}
Our framework's primary limitation is the computational cost of the Non-GAM approach. Future research should focus on performance optimization, such as developing specialized approximation techniques or leveraging hardware acceleration. Other promising avenues include: (1) broader empirical validation across diverse industrial datasets (e.g., e-commerce, finance), and (2) extending the model to handle continuous time-windows or integrate with causal inference methods to bridge the gap between mathematical and causal attribution.

\section{Ethical Considerations}
Our framework promotes transparency but requires responsible use. While it can uncover statistical bias, its fairness is contingent on the input data. The most significant risk is misinterpreting mathematical attribution as real-world causation; our method explains a numerical decomposition, not a causal story. This distinction is critical to prevent generating biased narratives or unfairly assigning blame. Finally, the Non-GAM approach necessitates that raw data be properly anonymized beforehand to protect privacy. Ultimately, our framework is a tool to augment, not replace, critical human judgment.


\bibliographystyle{plain} 
\bibliography{refs} 


\newpage
\appendix

\end{document}